\documentclass[twocolumn]{aastex631}
\usepackage{multirow}
\usepackage{longtable}

\shorttitle{Cyanopolyyne in L1544}
\shortauthors{Bianchi et al.}
\graphicspath{{./}{figures/}}

\begin{document}

\title{Cyanopolyyne chemistry in the L1544 prestellar core: new insights from GBT observations}

\correspondingauthor{Eleonora Bianchi}
\email{eleonora.bianchi@origins-cluster.de}

\author[0000-0001-9249-7082]{Eleonora Bianchi}
\affiliation{Excellence Cluster ORIGINS, Boltzmannstraße 2, D-85748 Garching bei München, Germany}
\affiliation{Ludwig-Maximilians-Universität, Schellingstraße 4, 80799 München, Germany}
\affiliation{Univ. Grenoble Alpes, CNRS, IPAG, 38000 Grenoble, France}
\affiliation{INAF, Osservatorio Astrofisico di Arcetri, Largo E. Fermi 5, I-50125, Firenze, Italy}

\author[0000-0001-9479-9287]{Anthony Remijan}
\affiliation{National Radio Astronomy Observatory, Charlottesville, VA 22903, USA}

\author[0000-0003-1514-3074]{Claudio Codella}
\affiliation{INAF, Osservatorio Astrofisico di Arcetri, Largo E. Fermi 5, I-50125, Firenze, Italy}
\affiliation{Univ. Grenoble Alpes, CNRS, IPAG, 38000 Grenoble, France}

\author[0000-0001-9664-6292]{Cecilia Ceccarelli}
\affiliation{Univ. Grenoble Alpes, CNRS, IPAG, 38000 Grenoble, France}

\author[0000-0002-0664-2536]{Francois Lique}
\affiliation{Univ. Rennes, CNRS, IPR [(Institut de Physique de Rennes)] - UMR 6251, F-35000 Rennes, France}

\author[0000-0002-6787-5245]{Silvia Spezzano}
\affiliation{Max-Planck-Institut f\"ur extraterrestrische Physik (MPE), 
Giessenbachstrasse 1, 85748 Garching, Germany}

\author[0000-0001-5121-5683]{Nadia Balucani}
\affiliation{Dipartimento di Chimica, Biologia e Biotecnologie, Via Elce di Sotto 8, 06123 Perugia, Italy}
\affiliation{INAF, Osservatorio Astrofisico di Arcetri, Largo E. Fermi 5, I-50125, Firenze, Italy}

\author[0000-0003-1481-7911]{Paola Caselli}
\affiliation{Max-Planck-Institut f\"ur extraterrestrische Physik (MPE), Giessenbachstrasse 1, 85748 Garching, Germany}

\author[0000-0002-4649-2536]{Eric Herbst}
\affiliation{Department of Chemistry, University of Virginia, Charlottesville, VA 22904, USA}
\affiliation{Department of Astronomy, University of Virginia, Charlottesville, VA 22904, USA}

\author[0000-0003-2733-5372]{Linda Podio}
\affiliation{INAF, Osservatorio Astrofisico di Arcetri, Largo E. Fermi 5, I-50125, Firenze, Italy}

\author[0000-0001-8211-6469]{Charlotte Vastel}
\affiliation{IRAP, Universit\'e de Toulouse, CNRS, UPS, CNES, 31400 Toulouse, France}

\author[0000-0003-1254-4817]{Brett McGuire}
\affiliation{Department of Chemistry, Massachusetts Institute of Technology, Cambridge, MA 02139, USA}
\affiliation{National Radio Astronomy Observatory, Charlottesville, VA 22903, USA}

\begin{abstract}

We report a comprehensive study of the cyanopolyyne chemistry in the prototypical prestellar core L1544.  Using the 100m Robert C. Byrd Green Bank Telescope (GBT) we observe 3 emission lines of HC$_3$N, 9 lines of HC$_5$N, 5 lines of HC$_7$N, and 9 lines of HC$_9$N.
HC$_9$N is detected for the first time towards the source. The high spectral resolution ($\sim$ 0.05 km s$^{-1}$) reveals double-peak spectral line profiles with the redshifted peak a factor 3-5 brighter. Resolved maps of the core in other molecular tracers indicates that the southern region is redshifted. Therefore, the bulk of the cyanopolyyne emission is likely associated with the southern region of the core, where free carbon atoms are available to form long chains, thanks to the more efficient illumination of the interstellar field radiation.
 We perform a simultaneous modelling of the HC$_5$N, HC$_7$N, and HC$_9$N lines, to investigate the origin of the emission. To enable this analysis, we performed new calculation of the collisional coefficients. The simultaneous fitting indicates a gas kinetic temperature of 5--12 K, a source size of 80$\arcsec$, and a gas density larger than 100 cm$^{-3}$. The HC$_5$N:HC$_7$N:HC$_9$N abundance ratios
measured in L1544 are about 1:6:4.
We compare our observations with those towards the the well-studied starless core TMC-1 and with the available measurements in different star-forming regions. The comparison suggests that a complex carbon chain chemistry is active in other sources and it is related to the presence of free gaseous carbon. Finally, we discuss the possible formation and destruction routes in the light of the new observations.

\end{abstract}


\section{Introduction} \label{sec:intro}

The formation of a Solar-type planetary system starts with the collapse of a cold ($\leq$10 K) and dense ($\geq$10$^{5}$ cm$^{-3}$) core, called a prestellar core, in a molecular cloud. The evolution of the prestellar core into a protostar, a protoplanetary disk and, eventually, a planetary system, is also accompanied by the evolution of its chemical composition (e.g. \citealt{Caselli2012}).


Cyanopolyynes are a class of molecules composed of a long chain of carbon atoms with a hydrogen atom at one end and a cyanide (CN) group at the other end. They are widespread in the interstellar medium (ISM) and they have been detected at all the stages of the star formation process, from dark clouds \citep[e.g.][]{Walmsley1980} to protoplanetary disks \citep{Chapillon2012}, and comets \citep[e.g.][]{Bock-Morvan2000}. While small cyanopolyynes such as HC$_{3}$N and HC$_{5}$N are regularly observed with (sub-)millimeter telescopes, much less is known about the presence and evolution of heavy species (e.g. chains with more than seven carbon atoms) which have their peak of emission at longer wavelengths. Even the relatively simple and abundant cyanotriacetylene (HC$_{7}$N) has only been detected in a handful of Solar-type prestellar cores and protostars (e.g. \citealt{Cernicharo1986,Gupta2009, Cordiner2012, Friesen2013, Jaber2017}). Yet, large carbon chains might have a crucial role in the heritage of organic material from the pre- and proto- stellar phase to the objects of the newly formed planetary system, such as asteroids and comets (e.g. \citealt{Mumma2011}).

Despite the importance of large carbon species in the astrobiological context and its potential diagnostic power, only the starless core TMC-1 has been extensively explored so far \citep[e.g.][]{Cernicharo2021a}.
This object has been the target of two deep surveys:  GOTHAM \citep[GBT Observations of TMC-1: Hunting Aromatic Molecules,][]{McGuire2020} and the QUIJOTE \citep[Q-band Ultrasensitive Inspection Journey to the Obscure TMC-1 Environment,][]{Cernicharo2021a} projects, which extensively investigated the cyanopolyyne chemistry in this source. In particular, they revealed for the first time several cyanopolyynes isotopologues and isomers (such as HC$_4$NC and HC$_6$NC) \citep{Cernicharo2020}, as well as the presence of HC$_{11}$N \citep{Loomis2021}, the largest cyanopolyyne so far discovered in the ISM.
Yet, TMC-1 actually is a starless core which does not have any sign of collapsing and becoming eventually a planetary system.
In this respect, the study of large cyanopolyynes in a prestellar core, which is believed to eventually form a Solar-type planetary system, is particularly important.

L1544, in the Taurus molecular cloud complex at a distance of 170 pc \citep[e.g.][]{Galli2019}, is considered the prototype of prestellar cores, being on the verge of gravitational collapse (e.g. \citealt{Caselli2012}). 
Its central high density ($\sim$10$^{6}$ cm$^{-3}$) and very low temperature ($\sim$ 7 K) result in the peculiar chemistry typical of cold and CO depleted gas, namely a very high deuteration of species \citep[e.g.][]{Caselli1999,Crapsi2007,Ceccarelli2007,caselli2022}. 
In the external layers, however, different rich chemical processes take place which lead to the formation of interstellar Complex Organic Molecules (iCOMs) and carbon chains \citep[e.g.][]{Bizzocchi2014, Vastel2016,Jimenez2016, Punanova2018,Ceccarelli2022}. 
Indeed, recent (single-dish) IRAM 30m observations in the mm-window show the presence of small carbon chains such as HC$_{3}$N, c-C$_{3}$H$_{2}$, C$_{3}$H, C$_{4}$H, C$_{2}$O and C$_{3}$O over extended portions of L1544 \citep{Vastel2014, Spezzano2017, Urso2019}. 

We carried out a pilot line survey in the X (8.0-11.6 GHz) and Ku (13.5-14.4 GHz) bands with the GBT towards L1544.
The goal of the project is to obtain the first census of molecular lines in this radio spectral range towards a well known and established analogue of the Solar System precursor.
In this first article, we report the study of the cyanopolyynes.

The article is organised as follows.
We report the observations and the detected lines in Sec. \ref{sec:obs-results}.
In Sec. \ref{sec:modeling}, we analyse the detected cyanopolyyne lines to derive constraints on the physical conditions of the emitting gas.
To this end, we carried out new computations to derive the collisional coefficients of HC$_5$N and HC$_7$N with H$_2$ in Sec. \ref{sec:nonLTE-coll-coef}.
Section \ref{sec:discussion} reports the implications of our new observation for the understanding of the cyanopolyyne chemistry in the earliest phases of a Solar-type planetary system and Sec. \ref{sec:conclusions} concludes the article.

\section{Observations and results} \label{sec:obs-results}

\subsection{Observations} \label{subsec:observations}

The observations presented here were carried out between 2019 June and 2020 February, on the GBT, under project codes AGBT19A$\_$048 and AGBT20A$\_$135.
The target source L1544 was observed at the coordinates $\alpha_{\rm J2000}$ = 05$^{\rm h}$ 04$^{\rm m}$ 16$\fs$60,
$\delta_{\rm J2000}$ = +25$\degr$ 10$\arcmin$ 48$\farcs$0.
The source calibrator 0530+1331 was used to perform pointing and focus. Observation were performed using position switching mode with an ON-OFF throw position of 1$\degr$. Two receivers were used to cover the X-band (8.0-11.6 GHz) and the Ku-band (13.5-14.4 GHz) in combination with the VEGA spectrometer in high-resolution mode. The bandwidth per spectral window was of 187.5 MHz with 131,072 channels corresponding to a resolution of 1.4 kHz (0.05 km s$^{-1}$ at 9 GHz). 
The r.m.s. is typically $\sim$ 5 mK in a channel of $\sim$ 1.4 kHz.
The telescope half-power beam width (HPBW) varies between $\sim$ 54$\arcsec$ in Ku-band and 1.4$\arcmin$ in X-band corresponding to $\sim$ 9180 au and $\sim$ 14280 au at the source distance.
The calibration was performed using GBTIDL. The spectra are first inspected by eye and cleaned of any artifacts.
Successively, each scan is corrected for Doppler tracking and calibrated in flux using the internal noise diodes from the receiver. The calibrated scans for a single observing session are then noise-weighted averaged.
The baseline subtraction is performed by automatically identifying the line-free channels and performing a polynomial fit.
The spectra were finally corrected for the GBT telescope efficiencies\footnote{https://www.gb.nrao.edu/scienceDocs/GBTpg.pdf}.
Calibration uncertainties are estimated to be $\sim$ 20$\%$.

\subsection{Results} \label{subsec:results}

\begin{figure}[ht]
\includegraphics[scale=0.7]{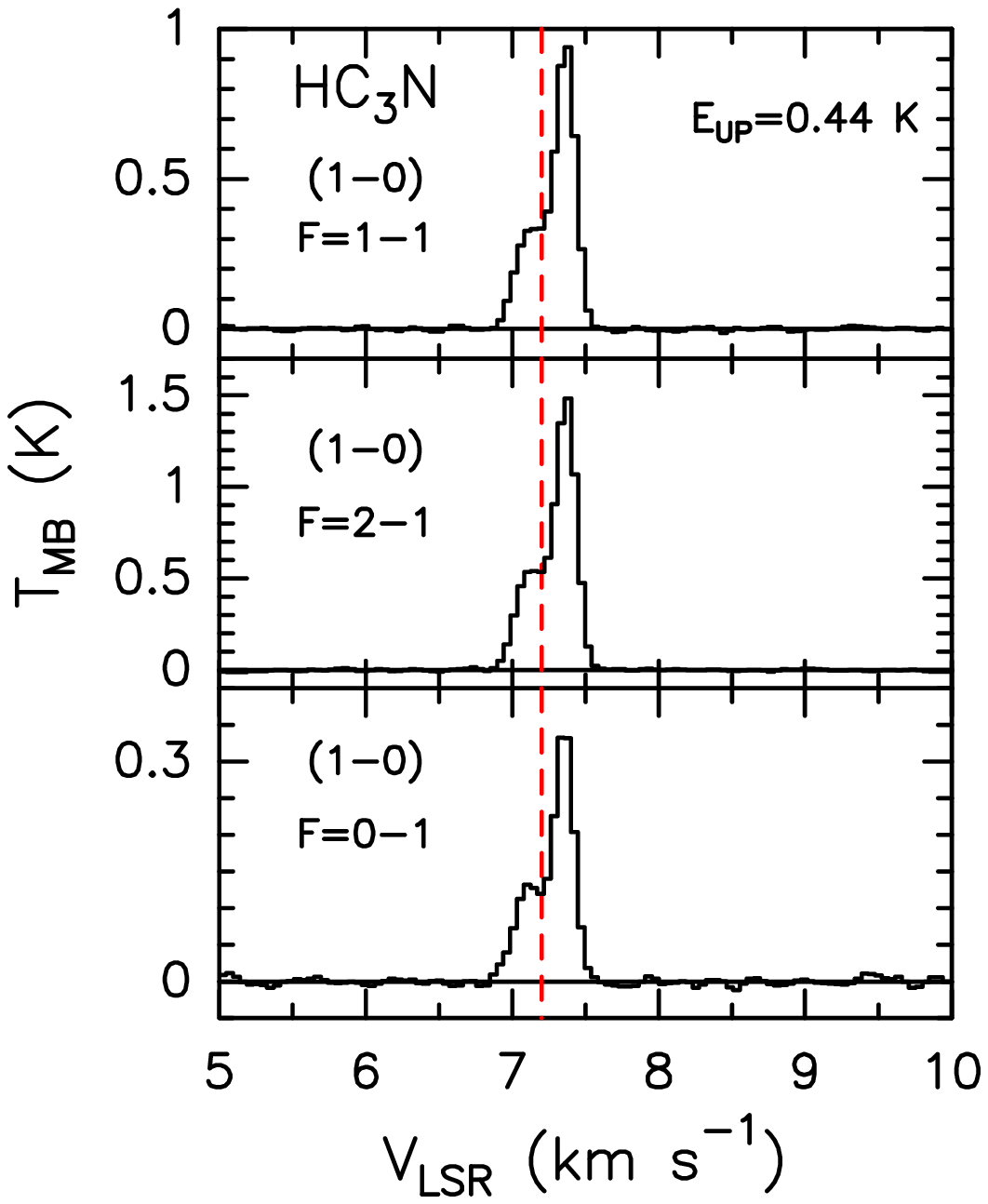}
  \caption{HC$_{\rm 3}$N transitions observed towards L1544 with the GBT. The vertical dashed lines mark the ambient LSR velocity (+ 7.2 km s$^{-1}$, \citealt{Tafalla1998}). The upper level energy of each transition is reported on the right inside the top panel.}
  \label{fig:spectra-HC3N}
\end{figure}

\begin{figure*}[ht]
\includegraphics[scale=0.7]{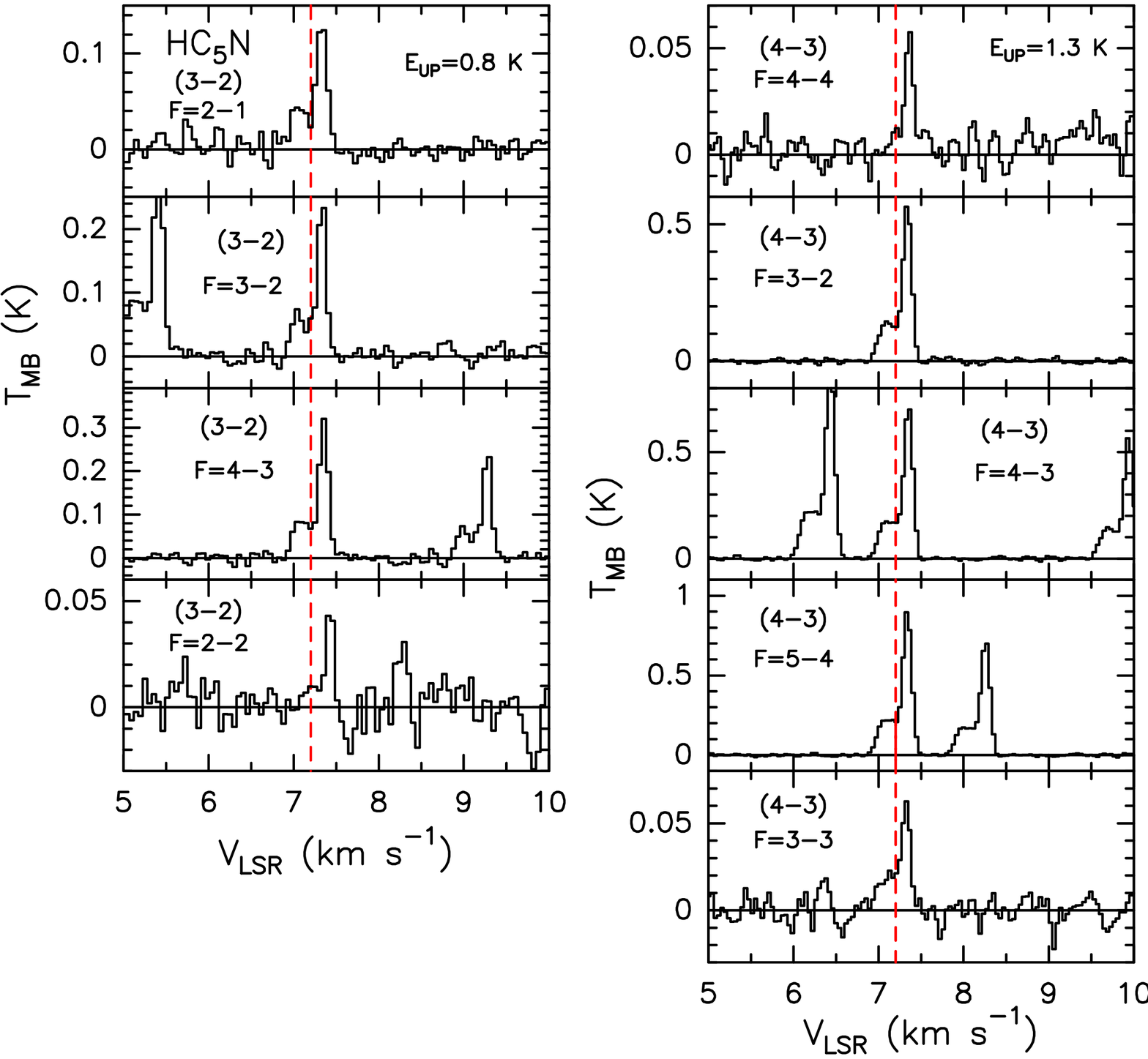}
  \caption{HC$_{\rm 5}$N transitions observed towards L1544 with the GBT. The vertical dashed lines mark the ambient LSR velocity (+ 7.2 km s$^{-1}$, \citealt{Tafalla1998}). The upper level energy of each transition is reported on the right inside the top panels.}
  \label{fig:spectra-HC5N}
\end{figure*}

\begin{figure}[ht]
\includegraphics[scale=0.7]{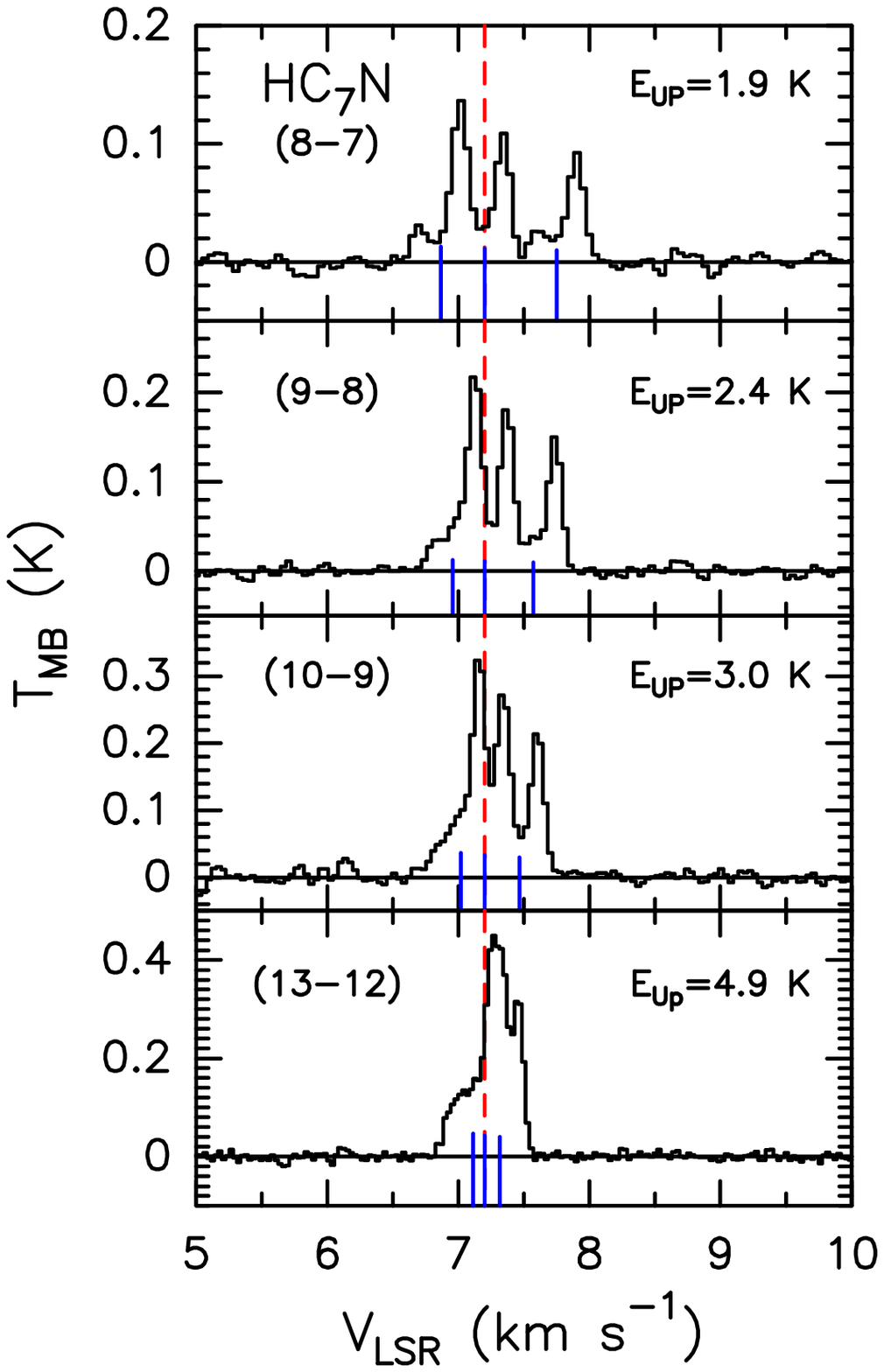}
  \caption{HC$_{\rm 7}$N transitions observed towards L1544 with the GBT. The vertical dashed lines mark the ambient LSR velocity (+ 7.2 km s$^{-1}$, \citealt{Tafalla1998}). The upper level energy of each transition is reported on the right inside each panel.}
  \label{fig:spectra-HC7N}
\end{figure}

\begin{figure*}[ht]
\includegraphics[scale=0.7]{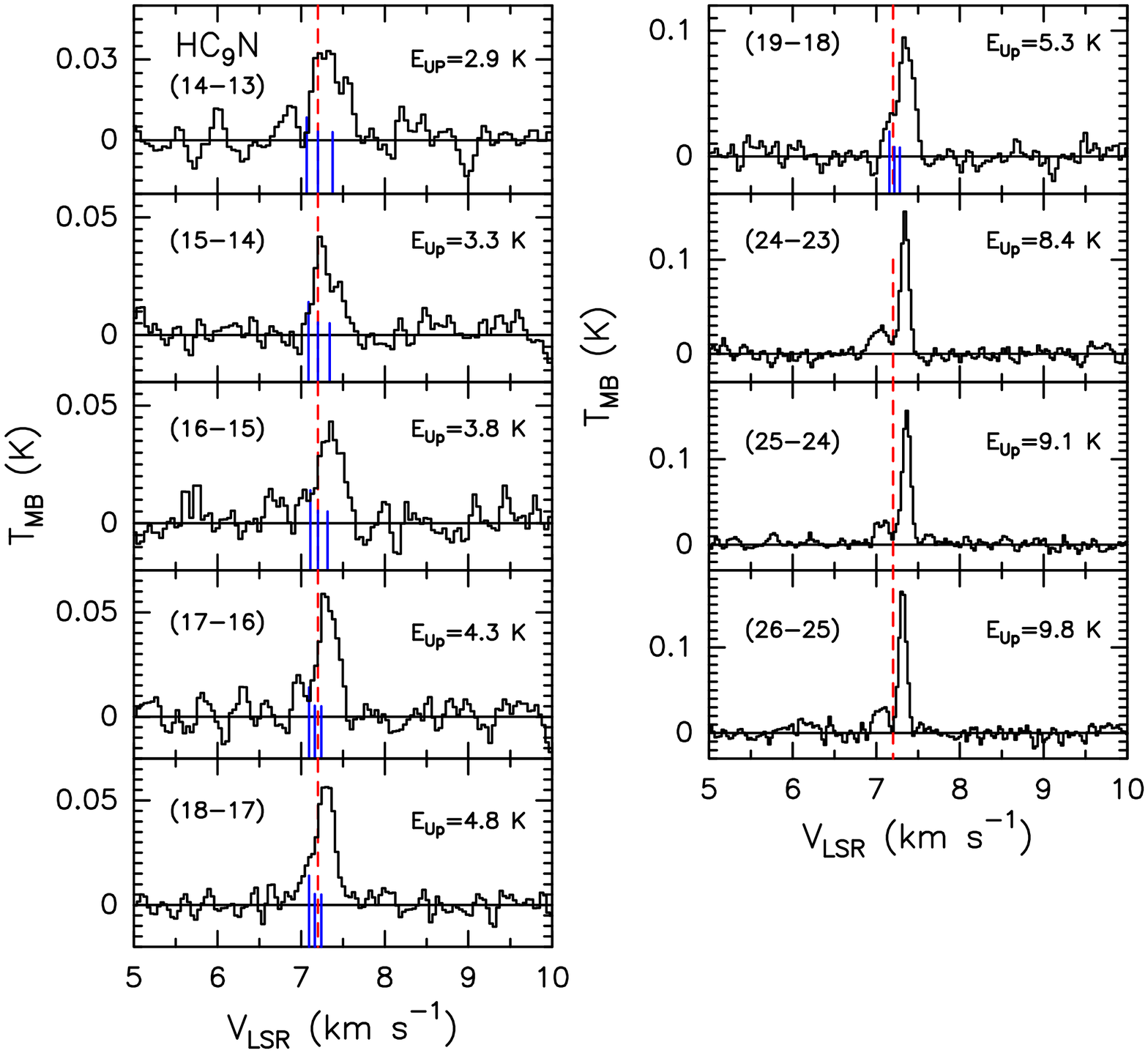}
  \caption{HC$_{\rm 9}$N transitions observed towards L1544 with the GBT. The vertical dashed lines mark the ambient LSR velocity (+ 7.2 km s$^{-1}$, \citealt{Tafalla1998}). The upper level energy of each transition is reported on the right inside each panel. The blue lines indicate the position of the hyperfine components.}
  \label{fig:spectra-HC9N}
\end{figure*}

We detected several bright emission lines from cyanopolyynes in L1544. More specifically, we detected 3 emission lines from HC$_3$N, 9 lines from HC$_5$N, 5 lines from HC$_7$N, and 9 lines from HC$_9$N (see Table \ref{Tab:lines}). The threshold for detection is a Signal-to-Noise larger than 3$\sigma$ at the line peak.
Line identification has been performed using the Cologne Database for Molecular Spectroscopy\footnote{\url{http://www.astro.uni-koeln.de/cdms/}} \citep{Muller2001, Muller2005}.
Table \ref{Tab:lines} reports the list of the detected transitions with their spectroscopic and derived line parameters, namely the line frequency, $\nu$, the upper level energy, E$_{\rm up}$, the line strength, S$\mu^2$, the line peak intensity in main beam temperature scale, T$_{\rm peak}$, the root mean square noise, rms, and the integrated intensity I$_{\rm int}$.
The upper level energy of the detected lines is low,
E$_{\rm up}$ $\leq$ 10 K.
The line analysis has been performed using the GILDAS\footnote{http://www.iram.fr/IRAMFR/GILDAS} CLASS package. The observed spectra are reported in Figures \ref{fig:spectra-HC3N} -- \ref{fig:spectra-HC9N}.

\startlongtable
\begin{deluxetable*}{lcccccc}
\tablecaption{List of transitions and line properties (in T$_{\rm MB}$ scale) of the cyanopolyyne emission. The columns report the transition and their frequency (GHz), the upper level energy E$_{\rm up}$ (K), the line strength $S\mu^2$ (D$^2$), the line peak temperature (mK), the rms (mK), and the velocity integrated line intensity I$_{\rm int}$ (mK km s$^{-1}$).
\label{Tab:lines}}
\tablewidth{0pt}
\tablehead{
\colhead{Transition} & \colhead{$\nu$$^{\rm a}$} & \colhead{E$_{\rm up}$$^{\rm a}$} & \colhead{$S\mu^2$$^{\rm a}$} & \colhead{T$_{\rm peak}$} & \colhead{rms} & \colhead{I$_{\rm int}$$^{\rm b}$} \\
\nocolhead{} & \colhead{(GHz)} & \colhead{(K)} & \colhead{(D$^2$)} &  \colhead{(mK)} & \colhead{(mK)} & \colhead{(mK km s$^{-1}$)}\\}
\startdata
\hline
\hline
\multicolumn7c{HC$_{\rm 3}$N} \\
\hline
HC$_{\rm 3}$N 1--0, F= 1--1 & 9.0970 & 0.44 & 4.6 & 941 & 5 &  247 (1) \\
HC$_{\rm 3}$N 1--0, F= 2--1 & 9.0983 & 0.44 & 7.7 & 1485 & 5 & 397 (1) \\
HC$_{\rm 3}$N 1--0, F= 0--1 & 9.1003 & 0.44 & 1.5  & 333 & 5 & 90 (1) \\
\hline
\multicolumn7c{HC$_{\rm 5}$N} \\
\hline
HC$_{\rm 5}$N 3--2, F= 2--1 &7.98778 & 0.77 & 33.8 & 121 & 8 & 30 (1)\\
HC$_{\rm 5}$N 3--2, F= 3--2 & 7.98799 & 0.77 & 50.0 & 229 & 8 & 47 (1)\\
HC$_{\rm 5}$N 3--2, F= 4--3 &7.98804 & 0.77 & 72.3 & 317 & 9 & 67 (1)\\
HC$_{\rm 5}$N 3--2, F= 2--2 &7.98992 & 0.77 & 6.2 & 43 & 6 & 7 (1)\\
HC$_{\rm 5}$N 4--3, F= 4--4 & 10.64923  & 1.28 & 4.7 & 54 & 6 & 9 (1)\\
HC$_{\rm 5}$N 4--3, F= 3--2 & 10.65056    & 1.28 & 53.6 & 567 & 6 & 108 (1)\\
HC$_{\rm 5}$N 4--3, F= 4--3 & 10.65065    & 1.28 & 70.3 & 700 & 7 & 141 (1) \\
HC$_{\rm 5}$N 4--3, F= 5--4 & 10.65068    & 1.28 & 91.7 & 896 & 6 & 179 (1)\\
HC$_{\rm 5}$N 4--3, F= 3--3 & 10.65249    & 1.28 & 4.7 & 61 & 6 & 13 (1)\\
\hline
\multicolumn7c{HC$_{\rm 7}$N} \\
\hline
\vspace{0.15cm}
%
HC$_{\rm 7}$N 8--7, F= 7--6 & 9.02399   & 	1.95 & 161.1 & 92 & 5 & 20 (1) \\
HC$_{\rm 7}$N 8--7, F= 8--7 & 9.02401  & \multirow{2}{*}{1.95} & 183.0 &  \multirow{2}{*}{109} & \multirow{2}{*}{5} & \multirow{2}{*}{46 (1)}\\
\vspace{1cm}
HC$_{\rm 7}$N 8--7, F= 9--8 &  9.02402 & & 207.7 & & & \\[+1cm]
HC$_{\rm 7}$N 9--8, F= 8--7  &    10.15200   & \multirow{3}{*}{2.44 }& 184.5 & \multirow{3}{*}{217} & \multirow{3}{*}{4} & \multirow{3}{*}{93 (1)} \\
HC$_{\rm 7}$N 9--8, F= 9--8 &  10.15201 & & 206.5& & & \\
\vspace{0.15cm}
HC$_{\rm 7}$N 9--8, F= 10--9 &  10.15202 & & 231.1& & & \\
HC$_{\rm 7}$N 10--9, F= 10--10  &  11.28900 & \multirow{3}{*}{3.00 }& 207.9 & \multirow{3}{*}{323} &  \multirow{3}{*}{8} & \multirow{3}{*}{133 (2)}   \\
HC$_{\rm 7}$N 10--9, F= 10--9 &  11.28001 & & 230.0 & & & \\
\vspace{0.15cm}
HC$_{\rm 7}$N 10--9, F= 11--10 & 11.28001 & & 254.4 & & &\\
HC$_{\rm 7}$N 13--12, F= 12--11  & 14.66399  & \multirow{3}{*}{4.93 }& 277.8 & \multirow{3}{*}{443} & \multirow{3}{*}{7} & \multirow{3}{*}{151 (1)} \\
HC$_{\rm 7}$N 13--12, F= 13--12 & 14.66399 & & 300.3 & & & \\
HC$_{\rm 7}$N 13--12, F= 14--13 & 14.66400 & & 324.4 & & & \\
\hline
\multicolumn7c{HC$_{\rm 9}$N} \\
\hline
HC$_{\rm 9}$N 14--13, F= 13--12  & 8.13450  & \multirow{3}{*}{2.93 }& 350.5 & \multirow{3}{*}{32} & \multirow{3}{*}{5} &\multirow{3}{*}{14 (1)} \\
HC$_{\rm 9}$N 14--13, F= 14--13 & 8.13450   & & 376.7 & & &  \\
\vspace{0.15cm}
HC$_{\rm 9}$N 14--13, F= 15--14 & 8.13451 & & 	404.6 & & & \\
HC$_{\rm 9}$N 15--14, F= 14--13  & 8.71553  & \multirow{3}{*}{3.35 }& 377.6 & \multirow{3}{*}{41}& \multirow{3}{*}{5}& \multirow{3}{*}{13 (1)} \\
HC$_{\rm 9}$N 15--14, F= 15--14 & 8.71554   & & 376.7 & & &   \\
\vspace{0.15cm}
HC$_{\rm 9}$N 15--14, F= 16--15 &8.71554 & & 	431.7 & & &  \\
HC$_{\rm 9}$N 16--15, F= 15--14  & 9.29657     & \multirow{3}{*}{3.79 }& 404.8 & \multirow{3}{*}{42} & \multirow{3}{*}{5} & \multirow{3}{*}{15 (1) } \\
HC$_{\rm 9}$N 16--15, F= 16--15 & 9.29657     & & 430.9  & & & \\
\vspace{0.15cm}
HC$_{\rm 9}$N 16--15, F= 17--16 & 9.29657   & & 458.9  & & & \\
HC$_{\rm 9}$N 17--16, F= 16--15  & 9.87760 & \multirow{3}{*}{4.27 }& 431.8 & \multirow{3}{*}{57} & \multirow{3}{*}{5} & \multirow{3}{*}{19 (1) } \\
HC$_{\rm 9}$N 17--16, F= 17--16 & 9.87761     & & 458.1 & \\
\vspace{0.15cm}
HC$_{\rm 9}$N 17--16, F= 18--17 & 9.87761  & & 486.0 & & & \\
HC$_{\rm 9}$N 18--17, F= 17--16  & 10.45864  & \multirow{3}{*}{4.77 }& 458.9 & \multirow{3}{*}{55}& \multirow{3}{*}{5}&\multirow{3}{*}{18 (1) } \\
HC$_{\rm 9}$N 18--17, F= 18--17 & 10.45864   & & 485.2 & & & \\
\vspace{0.15cm}
HC$_{\rm 9}$N 18--17, F= 19--18 & 10.45864     & & 513.0 \\
HC$_{\rm 9}$N 19--18, F= 18--17  & 11.03967  & \multirow{3}{*}{5.30 }& 486.0 & \multirow{3}{*}{95} & \multirow{3}{*}{7} &\multirow{3}{*}{22 (1) } \\
HC$_{\rm 9}$N 19--18, F= 19--18 & 11.03967       & & 512.3 & & &  \\
\vspace{0.15cm}
HC$_{\rm 9}$N 19--18, F= 20--19   & 11.03967     & & 540.1 & & & \\
HC$_{\rm 9}$N 24--23 & 13.94483  & 8.37 &1947.0 & 152 & 6 & 24 (1)  \\
HC$_{\rm 9}$N 25--24 & 14.52586 & 9.06 & 2028.1 & 157 & 6 & 25 (1) \\
HC$_{\rm 9}$N 26--25 & 15.10689 & 9.79 & 2108.9 & 166 & 7 & 26 (1) \\
\hline
\enddata
\tablecomments{$^{\rm a}$ Frequencies and spectroscopic parameters have been provided by \citet{deZafra1971} for HC$_{\rm 3}$N, \citet{Giesen2020} for HC$_{\rm 5}$N, and HC$_{\rm 9}$N, and \citet{McCarthy2000} for HC$_{\rm 9}$N, respectively, and retrieved from the Cologne Database for Molecular Spectroscopy \citep{Muller2005}.
$^{\rm b}$ Errors on the integrated intensity do not include 20$\%$ of calibration.\\
}
\end{deluxetable*}






The observed cyanopolyyne lines show a double peaked profile, revealed thanks to the very high spectral resolution ($\sim$ 1.4 kHz, corresponding to $\sim$ 30 m s$^{-1}$ at 14 GHz) provided by the GBT.
The two emission peaks are located at +7.1 km s$^{-1}$ and +7.3 km s$^{-1}$ with a dip located at the systemic source velocity (+7.2 km s$^{-1}$). The red-shifted peak is brighter than the blue-shifted one by a factor between 3 and 5, depending on the line.
We report for the first time towards the source the heavier cyanopolyyne HC$_9$N, while HC$_3$N, HC$_5$N, and HC$_7$N have been previously observed in L1544 \citep{Snell1981, Cernicharo1986,Quenard17, Hily2018}. The HC$_3$N spectra previously observed at the IRAM 30m have a spectral resolution $\sim$ 0.2 km s$^{-1}$, and they do not reveal the double peak profile.
Interestingly, the velocities of the two peaks are consistent with the blue- and red-shifted velocities
revealed by the peak velocity distribution derived by \citet{Spezzano2016} in c-C$_3$H$_2$ using 
the IRAM 30m. 
In addition, the spectral profiles of the present data set are well consistent with those at high spectral resolution (down to $\simeq$ 0.04 km s$^{-1}$)
previously observed at GBT towards L1544 in other complex C-bearing species such as C$_4$H, C$_6$H and C$_6$H$^{-}$ \citep{Gupta2009}.

\section{Physical conditions and abundance ratios} \label{sec:modeling}

\subsection{Full radiative transfer modeling}\label{subsec:model-loc}

\begin{figure*}[ht]
\includegraphics[scale=0.5]{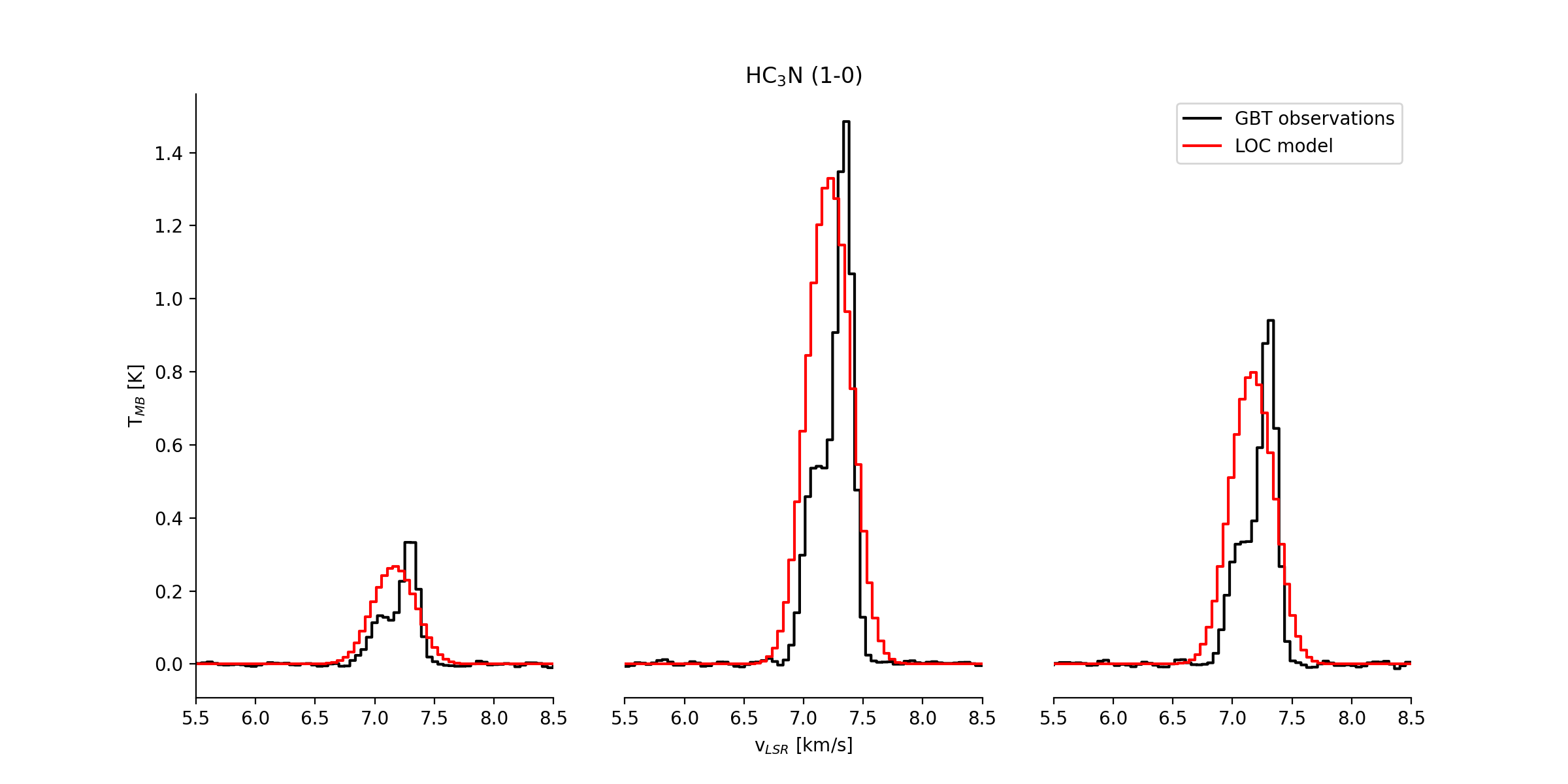}
  \caption{Comparison between the observed HC$_3$N line profiles and the theoretical profiles predicted using the radiative transfer code LOC (see Sect. \ref{subsec:model-loc}). The model, assuming spherical symmetry distribution, reproduce the line intensities but not the shape of the profiles. This suggests an asymmetric distribution of cyanopolyynes across the core, with the red-shifted component brighter than the blue-shifted one.}
  \label{fig:LOC-HC3N}
\end{figure*}

In order to interpret the line profiles, we use the full radiative transfer code LOC \citep{Juvela2020} with a 1D model that assumes spherically symmetric distribution of physical structures, characterised by volume density, $\rho(r)$, kinetic temperature, $T(r)$ and velocity field, including both micro-turbulence, $\sigma_{\mathrm{turb}}$ and radial velocity, $V(r)$.  
We adopted the parameterised forms of $\rho(r)$, $T(r)$ and $V(r)$ following \citet{Keto2010} for L1544. 
The core radius is assumed to be 0.3 pc. In the modeling, linear discretization is used for the grids with a physical resolution of 60 au. We then convolved the output spectral cube with the corresponding observational beam for each frequency. We tested several abundance profiles for HC$_3$N, with both constant abundance profile and constant abundance profile with depletion in the inner few thousand au of the core. While we are able to reproduce the overall intensity of the lines with abundances of few 10$^{-9}$, we cannot reproduce the observed line profiles.
Figure \ref{fig:LOC-HC3N} shows the results of one of our test, ran using a constant abundance profile of 5$\times$10$^{-9}$ with respect to H$_2$ and complete freeze-out towards the inner 5000 au of the core.
The 1-0  transition of HC$_3$N, given the very low critical density ($\sim$10$^3$ cm$^{-3}$), traces the outer layers of the core, as also found by e.g. \citet{Liu2022} in cold Planck clumps, and it is likely that the line profile that we observe is showing us the asymmetry in these outer layers. As a consequence, by using a spherical model we cannot reproduce the line profiles. The profile of many emission lines have been successfully reproduced with the Keto \& Caselli physical structure of L1544 using transitions with larger critical densities (10$^4$-10$^5$ cm$^{-3}$) with respect to the 1-0 transition of HC$_3$N. 
Likely the spherical symmetry approximation is not realistic for transitions tracing low density gas.



\subsection{Large Velocity Gradient modelling} \label{subsec:model-LVG}
In order to understand the nature and the spatial origin of the gas emitting in cyanopolyynes, we analysed the observed HC$_5$N, HC$_7$N and HC$_9$N lines via a non Local thermodynamic equilibrium (non-LTE) Large Velocity Gradient (LVG) approach.
To this end, we used the code \textsc{grelvg}, initially developed by \citet{Ceccarelli2003}.
We used the collisional coefficients of HC$_5$N, HC$_7$N and HC$_9$N with para-H$_2$ between 10 and 100 K, computed from the HC$_3$N collisional coefficients as described in Sec. \ref{sec:nonLTE-coll-coef}.
The first 50 levels of each species are included, which means transitions with upper level energies of 160, 70 and 35 K for HC$_5$N, HC$_7$N and HC$_9$N, respectively.
We will discuss the impact of these limits when analysing the results of the modeling.
To compute the line escape probability as a function of the line optical depth we adopted a semi-infinite expanding slab geometry \citep{Scoville1974} and a line width equal to 0.4 km~s$^{-1}$, following the observations.

We ran several grids of models to sample the $\chi^2$ surface in the parameters space.
Specifically, we varied the HC$_5$N column density N(HC$_5$N) from $1\times 10^{12}$ to $2\times 10^{15}$ cm$^{-2}$, the HC$_5$N/HC$_7$N abundance ratio $f_{5-7}$ from 3 to 9, the HC$_5$N/HC$_9$N abundance ratio $f_{5-9}$ from 9 to 49, the H$_2$ density n$_{H2}$ from 100 to $10^{6}$ cm$^{-3}$ and the temperature T$_{gas}$ from 5 to 100 K.
We then fitted the measured HC$_5$N, HC$_7$N and HC$_9$N velocity-integrated line intensities by comparing them with those predicted by the model, leaving N(HC$_5$N), $f_{5-7}$, $f_{5-9}$, n$_{H2}$, $T_{gas}$ and the emitting size $\theta$ as free parameters.

We proceeded in two steps.
In a first step, we considered the three species independently and obtained constraints on their column densities and the emitting gas properties ($\theta$, n$_{H2}$, T$_{gas}$).
In a second step, we fitted the lines from the three species simultaneously, obtaining constraints on all parameters: N(HC$_5$N), $f_{5-7}$, $f_{5-9}$, $\theta$, n$_{H2}$ and T$_{gas}$.

\subsubsection{Step 1: HC$_5$N, HC$_7$N and HC$_9$N separate fit}\label{subsubsec:Step1}

\textit{HC$_5$N:}
Given the limited number (two) of lines and range of upper level energy, only a lower limit to the N(HC$_5$N) is obtained, $\geq 4\times$ 10$^{13}$ cm$^{-2}$, with an emitting size of about 100$''$.

\vspace{0.2cm}
\textit{HC$_7$N:}
With four lines, the HC$_7$N fitting constrains better the gas parameter than the HC$_5$N one.
The column density N(HC$_7$N) is constrained at 1 $\sigma$ between 0.6 and 3 $\times$ 10$^{13}$ cm$^{-2}$, where the emitting size are from 100$''$ to 38$''$, respectively.
Both the gas temperature and densities are difficult to constrain as they depend on the HC$_7$N column density (see \S ~\ref{subsubsec:Step2} below).

\vspace{0.2cm}
\textit{HC$_9$N:}
The column density N(HC$_9$N) is, at 1 $\sigma$, between 0.1 and 3 $\times$ 10$^{13}$ cm$^{-2}$, where the emitting size are from 150$''$ to 38$''$, respectively.
Again, the gas temperature and densities are difficult to constraint as they depend on the HC$_9$N column density (see \S ~\ref{subsubsec:Step2} below).

\subsubsection{Step 2: HC$_5$N, HC$_7$N and HC$_9$N simultaneous fit}\label{subsubsec:Step2}
We assumed that the three cyanopolyynes originate in the same gas.
This assumption is based on the similarity of the line shapes of the three species.
We considered all the lines detected in the observations presented here, namely a total of 18 lines which leads to 12 degrees of freedom in the fit.

\begin{figure*}
    \centering
    \includegraphics[width=1\columnwidth,trim=0 0.5cm 0 1.2cm]{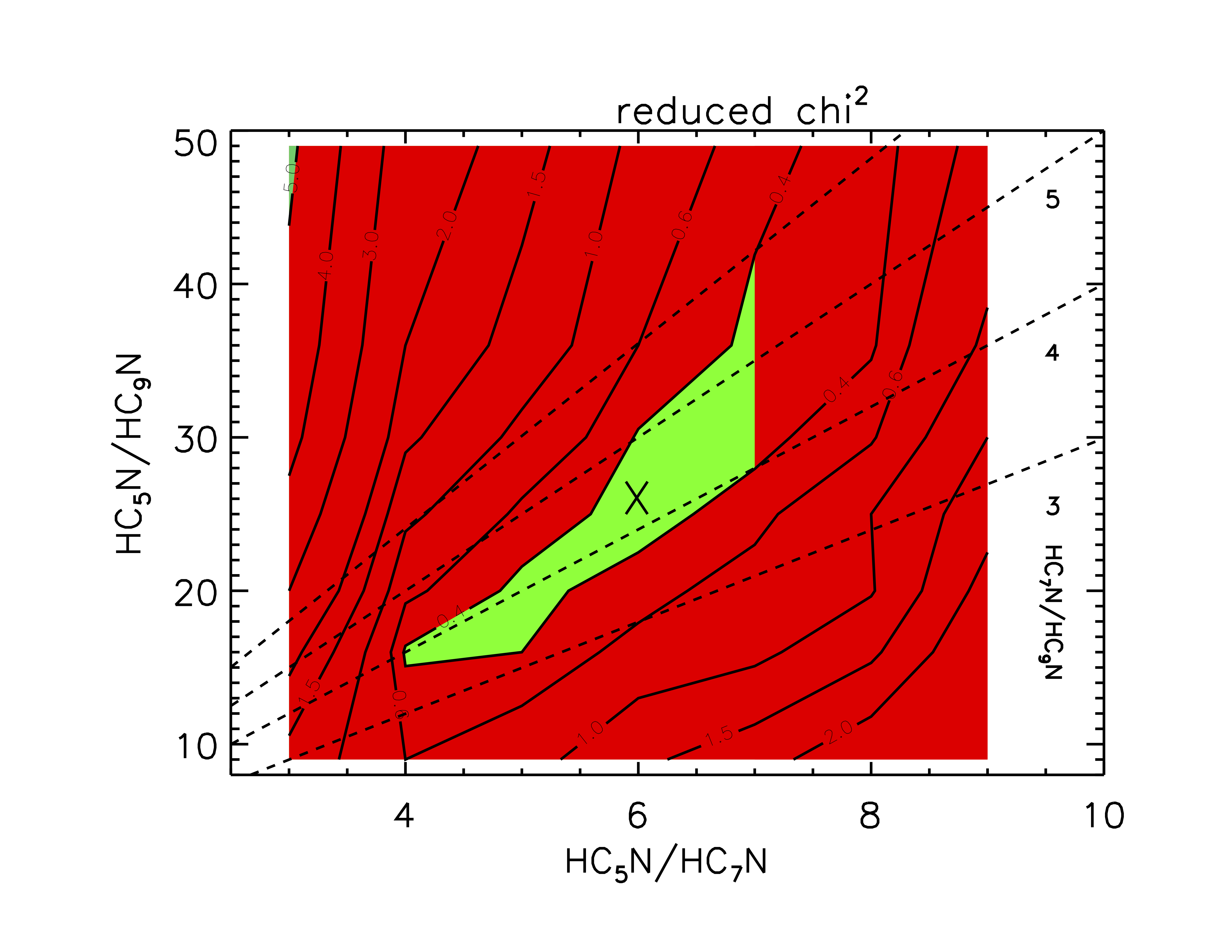}
    \includegraphics[width=1\columnwidth,trim=0 0.5cm 0 1.2cm]{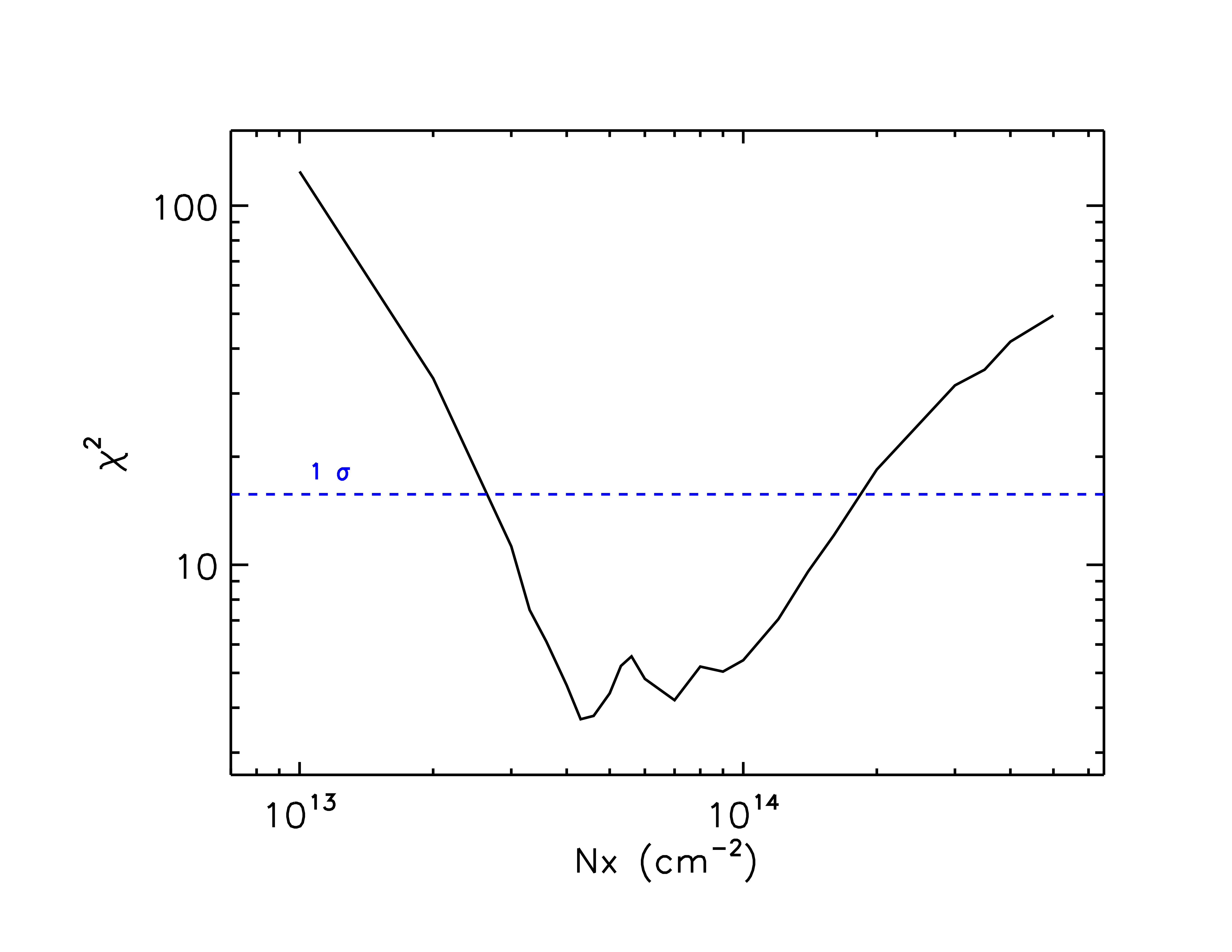}
    \includegraphics[width=1\columnwidth,trim=0 0.5cm 0 1.2cm]{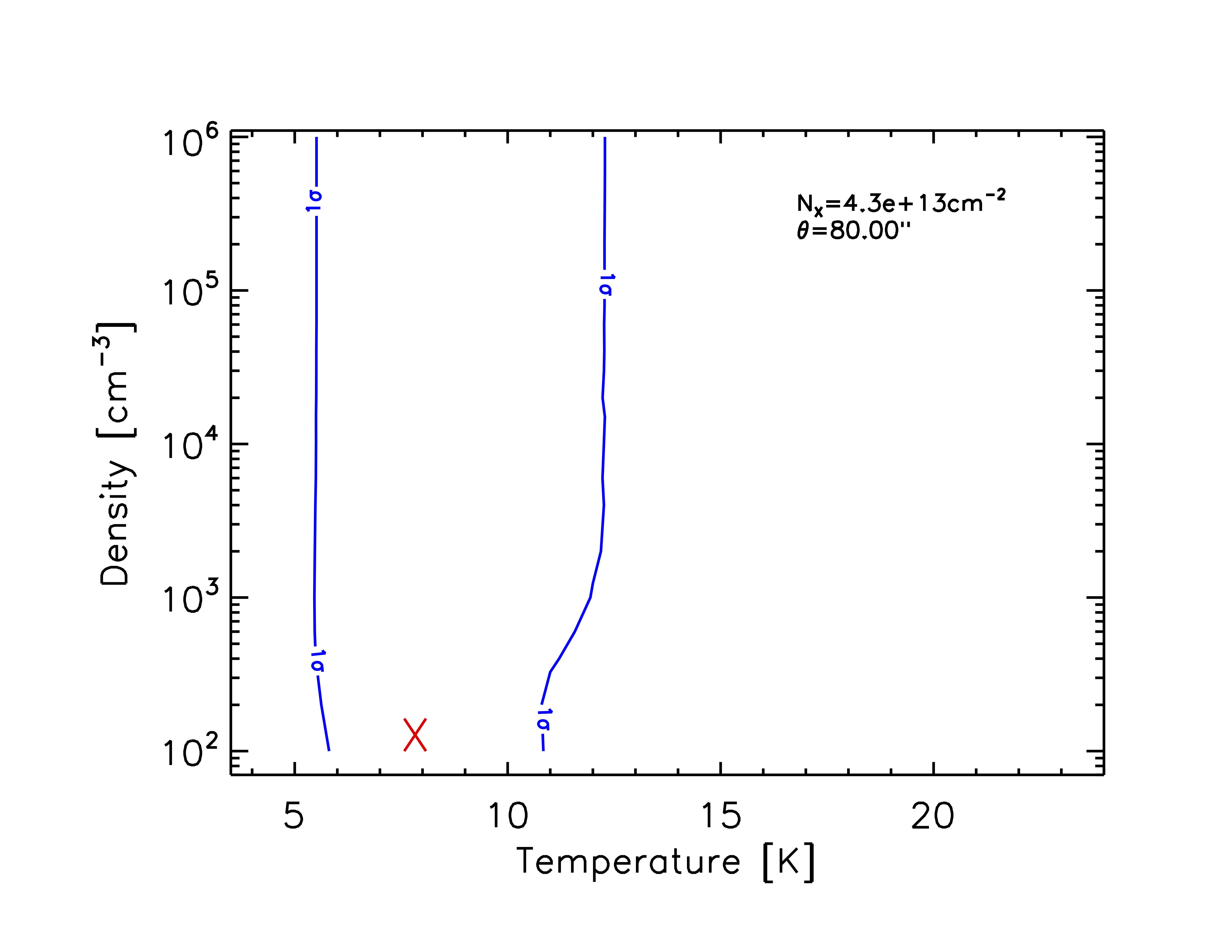}
    \includegraphics[width=1\columnwidth,trim=0 0.5cm 0 1.2cm]{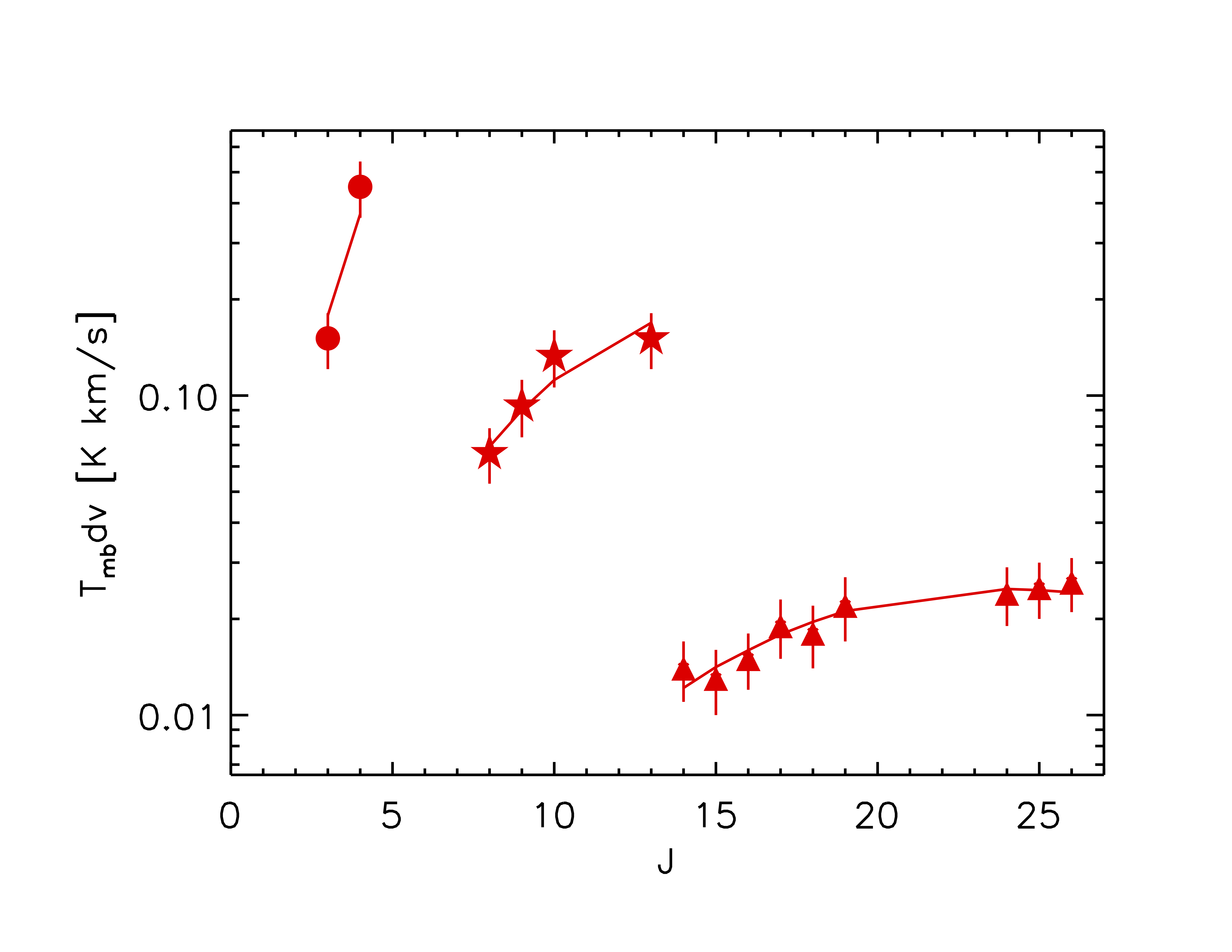}

    \caption{Results from the $\chi^2$ minimisation obtained fitting simultaneously the HC$_5$N, HC$_7$N and HC$_9$N velocity-integrated line intensities.
    \textit{Top left panel}: Reduced $\chi^2$ as a function of the HC$_5$N/HC$_7$N and HC$_5$N/HC$_9$N abundance ratios. 
    The cross shows the best fit and the green area the 1$\sigma$ uncertainty parameter space.
    The dashed lines show the HC$_7$N/HC$_9$N ratio.
    \textit{Top right panel}: Minimum $\chi^2$ obtained at each HC$_5$N column density N(HC$_5$N) (minimised with respect to the gas temperature and density) as a function of N(HC$_5$N).
    The dashed horizontal line shows the 1$\sigma$ interval.
    \textit{Bottom left panels}: Contour plot of the $\chi^2$ as a function of the density and temperature, obtained for the best fit HC$_5$N column density of $4.3\times$ 10$^{13}$ cm$^{-2}$ and source size of 80$^{\prime\prime}$ (see upper panel).
    The red cross indicates the best fit gas density and temperature and the blue curve the 1$\sigma$ interval.
    \textit{Bottom right panel}: Observed velocity-integrated line intensities of HC$_5$N (circles), HC$_7$N (stars) and HC$_9$N  (triangles) and the modelled ones (red curves) as a function of the upper J of the transitions, computed at the best fit (see text).}
    \label{fig:nonLTE-analysis}
\end{figure*}

We started by exploring the $\chi^2$ surface over large ranges and gradually zoomed to smaller ones to be sure not to miss local minima.
In practice, each grid consisted of about 10$^4$ models and we ran a dozen grids.
The results of the analysis are shown in Fig. \ref{fig:nonLTE-analysis} and reported in Table \ref{tab:LVG-results}.

\begin{table}[]
    \centering
    \begin{tabular}{c|c|c}
    \hline
         Parameter [units] & Best-fit & 1$\sigma$ range \\
         \hline
         $\theta$ [$^{\prime\prime}$] & 80 & 32--100 \\
         N(HC$_5$N) [$\times 10^{13}$ cm$^{-2}$]& 4.3 & 3--20 \\
         N(HC$_5$N)/N(HC$_7$N) & 6 & 4--7 \\
         N(HC$_7$N)/N(HC$_9$N) & 4 & 3--5 \\
         H$_2$ density [cm$^{-3}$] & 100 & $\geq 100$ \\
         Gas temperature [K] & 7.5 & 5--12\\ 
         \hline
    \end{tabular}
    \caption{Results of the non-LTE LVG analysis. 
    Best fit emitting region size (row 1), HC$_5$N column density (row 2), HC$_5$N/HC$_7$N and HC$_5$N/HC$_9$N column density ratios (row 3 nd 4, respectively), H$_2$ density (row 5), and gas temperature (row 6). For each parameter the 1$\sigma$ range is reported in column 2.}
    \label{tab:LVG-results}
\end{table}


First, the top panel of Fig. \ref{fig:nonLTE-analysis} shows the reduced-$\chi^2$ as a function of the HC$_5$N/HC$_7$N and HC$_5$N/HC$_9$N abundance ratios.
The best fit (reduced-$\chi^2$= 0.31) is obtained for $f_{5-7}=6^{+1}_{-2}$ and $f_{5-9}=25 \pm10$, along the a banana-like curve with HC$_7$N/HC$_9$N=$4\pm1$.
The best fit is obtained at HC$_5$N/HC$_7$N=6 and HC$_7$N/HC$_9$N=4.
With these values, the HC$_5$N column density N(HC$_5$N) is equal to $\sim$3--20 $\times 10^{13}$ cm$^{-2}$ and $\theta$ is $\sim$100--32$\arcsec$ (the larger N(HC$_5$N) the smaller $\theta$), as shown in Fig. \ref{fig:nonLTE-analysis}.
The lowest $\chi^2$ is obtained for N(HC$_5$N)=$4.3\times 10^{13}$ cm$^{-2}$ and $\theta$=80$\arcsec$.
With these values, the lowest $\chi^2$ is obtained for $T_{gas}$=7.5 K and n$_{H2}$=100 cm$^{-3}$.
At the 1$\sigma$ level the temperature is between 5 and 12 K and the density remains unconstrained (Fig. \ref{fig:nonLTE-analysis}).
The observed and predicted intensities of all the lines are shown in the bottom panel of Fig. \ref{fig:nonLTE-analysis}.
The HC$_5$N, HC$_7$N and HC$_9$N lines are all optically thin (the highest $\tau$ is 0.6 for the HC$_5$N J=4 line) and the line are moderately subthermally populated (T$_{ex} \sim$6--7 K).

\subsubsection{Impact of the limited number of considered levels}
Finally, we comment about the possible impact on the gas temperature caused by the limited number of levels, only 50 for the three cyanopolyynes.
While the number of levels of HC$_5$N are enough for a reasonable analysis for temperatures below 30 K, the limited number of levels of HC$_7$N and HC$_9$N may, in principle, be problematic.
However, since the gas temperature derived from the previous LVG analysis is lower than 12 K, the analysis is probably not greatly impacted by the range of energy levels probed by the observations.
For example, the excitation temperature T$_{ex}$ of the HC$_7$N \textit{J}=13-12 and HC$_9$N \textit{J}=26-25 lines are 7.3 and 6.95 K, respectively, when the gas kinetic temperature is 7.5 K. 
Therefore, the highest levels of HC$_7$N and HC$_9$N are probably not populated, which makes the above analysis reliable.

\section{Collisional coefficients}\label{sec:nonLTE-coll-coef}

To the best of our knowledge, no collisional data are available for the HC$_5$N and HC$_7$N molecules.
However, as already suggested by \cite{Snell:81}, HC$_5$N--H$_2$ and HC$_7$N--H$_2$ rate coefficients may be estimated from HCN--H$_2$ and HC$_3$N--H$_2$ rate coefficients considering that the rate coefficients will be proportional to the size of the molecules. 

We considered the HC$_3$N--H$_2$ rate coefficients  as a reference for the estimation of HC$_5$N--H$_2$ and HC$_7$N--H$_2$ rate coefficients. As a crude approximation, HC$_5$N and HC$_7$N are respectively 1.5 and 2 times longer than HC$_3$N whereas HCN is 2 times shorter than HC$_3$N.

\cite{Snell:81} scale the HC$_3$N rate coefficients by a factor 1.5 and 2 to get HC$_5$N and HC$_7$N rate coefficients respectively. They also checked that HCN and HC$_3$N rate coefficients obey to the same rules. However, as noticed by \cite{Snell:81}, the scaling factors are averages and significant deviations occur depending on the transition and the temperature considered.

In this work, in order to improve the accuracy of the estimation, we consider scaling factors depending on the size of the molecule (1.5 for HC$_5$N and 2 for HC$_7$N) but also on the transition and the temperature considered. Indeed, the ratio of HC$_3$N--H$_2$ over HCN--H$_2$ rate coefficients was also used to evaluate the HC$_5$N--H$_2$ and HC$_7$N--H$_2$ as follows:

\begin{eqnarray}
k^{\rm{HC_7N-H_2}}_{J \to J'} (T) & = & k^{\rm{HC_3N-H_2}}_{J \to J'} (T) \frac{k^{\rm{HC_3N-H_2}}_{J \to J'} (T)}{k^{\rm{HCN-H_2}}_{J \to J'} (T)}
\end{eqnarray}
and
\begin{eqnarray}
k^{\rm{HC_5N-H_2}}_{J \to J'} (T) & = &  \frac{1}{2} (k^{\rm{HC_3N-H_2}}_{J \to J'} (T)+k^{\rm{HC_7N-H_2}}_{J \to J'} (T))
\end{eqnarray}
The HCN--H$_2$ rate coefficients of \cite{benabdallah:12} and the HC$_3$N--H$_2$ rate coefficients of \cite{Wernli:07} were used in the above formula. 

Using such an approach, we expect to have a better description of large $\Delta J$ (\textit{J}'-\textit{J} $>>$ 1) for collisionally-induced transitions and for very low temperatures. In the astrophysical models, the HC$_9$N--H$_2$ rate coefficients were considered to be the same as the HC$_7$N--H$_2$ ones. It should be noted that the accuracy of the present calculations is expected to be accurate only to within about an order of magnitude. Future calculations of rate coefficients for HC$_5$N and HC$_7$N molecules have to be performed using a more reliable approach.

\section{Discussion} \label{sec:discussion}

\subsection{Origin of the cyanopolyyne emission in L1544} \label{subsec:discussion-origin}

The high spectral resolution provided by the present observations allow us to resolve in details the shape of the observed cyanopolyynes lines (see Sec \ref{subsec:results}).
Specifically, the lines present a double peak with the red-shifted intensity brighter than the blue-shifted one (Sec. \ref{subsec:results}).
Previous maps by \cite{Spezzano2016} and \cite{Spezzano2017} show that the emission of carbon chains is concentrated in south-east region of L1544, in contrast with that of methanol, whose lines are bright in the north.
The velocity in the two regions is slightly different, with the C-chain peak red-shifted with respect to the velocity of the methanol peak.
In the present observations, the velocity of the cyanopolyyne red-shifted peak is consistent to that of the carbon chains emitting region.
Therefore, the fact that the cyanopolyynes lines are brighter in the red-shifted peak implies that they mainly originate in the south-east region of the L1544 core.
These results support the idea that the carbon chains abundance is enhanced towards the external part of the core, where material is more exposed to the interstellar field radiation. 
The southern part of the source is particularly exposed since it is located at the edge of the cloud \citep{Andre2010}, and this would increase the free carbon atoms available to form long chains (see also \citealt{Spezzano2016}).
The line profiles observed with the GBT are consistent with those observed by \citet{Gupta2009} in other carbon chains but clearly different from those observed in high-density gas tracers towards L1544 (e.g., N$_2$D$^+$ and N$_2$D$^+$; \citealt{Caselli2002,Caselli2017}), supporting this interpretation. 

The non-LTE LVG analysis (Sec. \ref{subsec:model-LVG}) indicates that the cyanopolyyne emission originates predominantly from an extended region ($\sim$ 13600 au in radius) at low temperature (5--12 K). 
Unfortunately, the gas density is unconstrained other than being larger than $\sim$ 100 cm$^{-3}$.

\begin{figure}
    \centering
    \includegraphics[width=1\columnwidth]{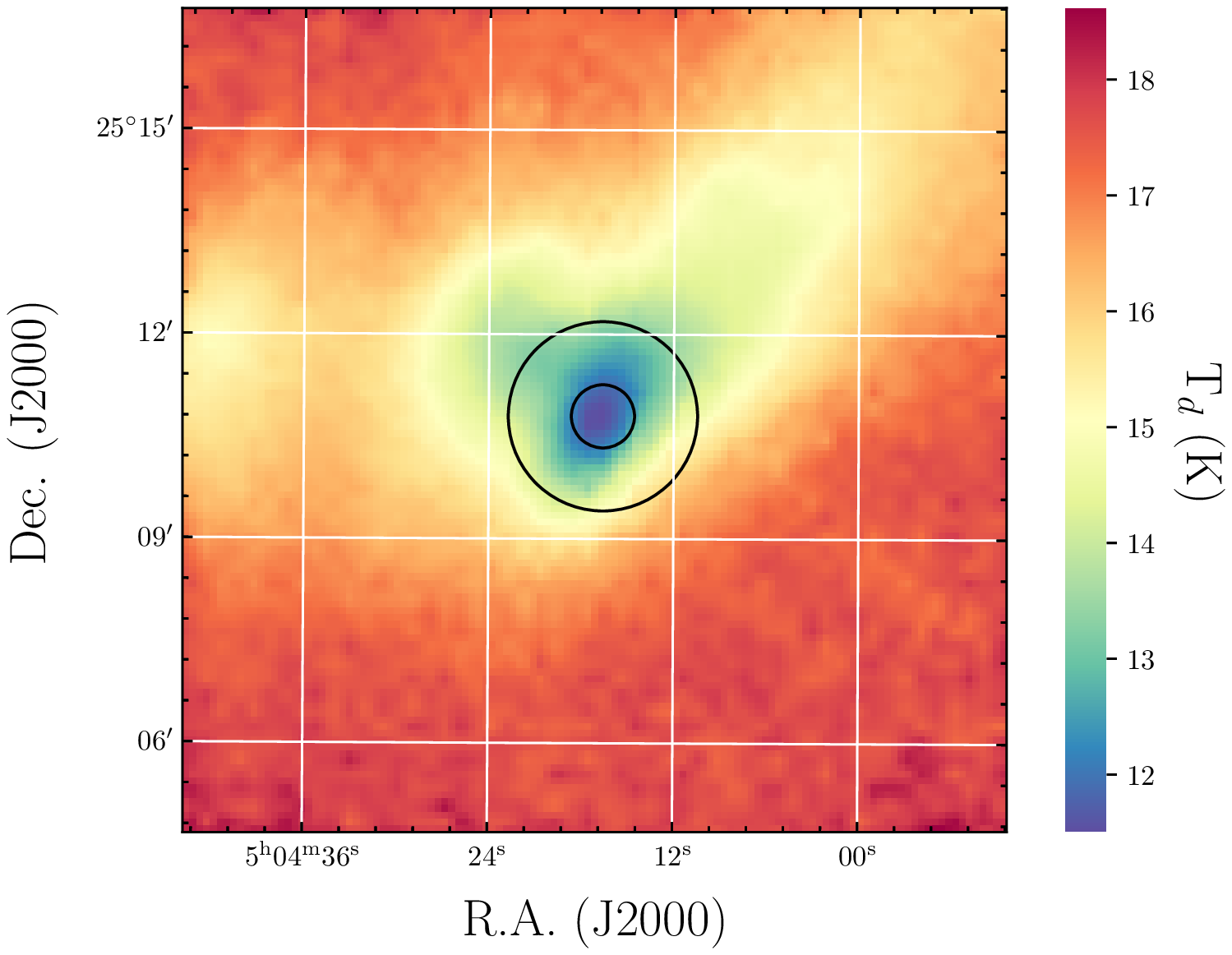}
    \caption{Dust temperature calculated from l/SPIRE data at 250, 350, and 500 $\mu$m observations towards L1544 and presented in \citet[][]{Spezzano2016}. We show superposed in black the GBT HPBWs in the two observed bands.}
    \label{fig:herschel}
\end{figure}

The derived temperature is in agreement with the dust temperature measured with Herschel and shown in Fig. \ref{fig:herschel}. SPIRE is only sensitive to the extended emission and the core regions probed by the GBT have a dust temperature ranging between 11.5 K and 15 K. On the other hand, other molecular tracers, such as NH$_3$ and its isotopologues, show that the gas temperature decreases towards the central part of the core down to $\sim$ 7 K \citep{Crapsi2007}.

However, the analysis has several limitation which should be taken into account. 
First, the physical model of the source used in our analysis of Sec. \ref{subsec:model-loc}, i.e. the one by \citet{Keto2010}, assumes spherical symmetry, within the observed region. On the contrary, the non-LTE LVG analysis in Sec. \ref{subsec:model-LVG}, assumes a semi-infinite slab.
On the other hand, the spectral profiles suggest that the cyanopolyyne emission is non-homogeneously distributed and the modelling does not distinguish the different emitting components.
The results of the modelling would be then representative of the physical conditions of the main emitting component but they could be locally different. 
Another assumption of the Sec. \ref{subsec:model-LVG} non-LTE LVG modelling is that all the cyanopolyynes are cospatial and trace the same gas. 
Although this hypothesis is supported by the similar line profiles, it can not be verified only using single-dish GBT observations. 
Finally, recent observations have shown the presence of deuterated carbon chains (c-C$_3$HD and c-C$_3$D$_2$) towards the dusty peak at the center of the core \citep{Giers2022}. 
This would suggest that even if the bulk of the emission comes from the external layers, a significant fraction of molecular species rich in carbon can still be present in the densest regions of the core, where free carbon atoms are produced by the CO destruction from the molecular ions created by cosmic-rays (such as He$^{+}$), as first proposed by \citet{Ruffle1999}.
Further interferometric observations are needed to map the cyanopolyyne distribution across the core. In this respect, the next generation of radio interferometer such as SKA$\footnote{https://www.skao.int/}$ and ngVLA$\footnote{https://ngvla.nrao.edu/}$ will be a major step ahead.

\subsection{Comparison of L1544 with other sources} \label{subsec:discussion-comparison}

\begin{figure*}
    \centering
    \includegraphics[width=1\columnwidth]{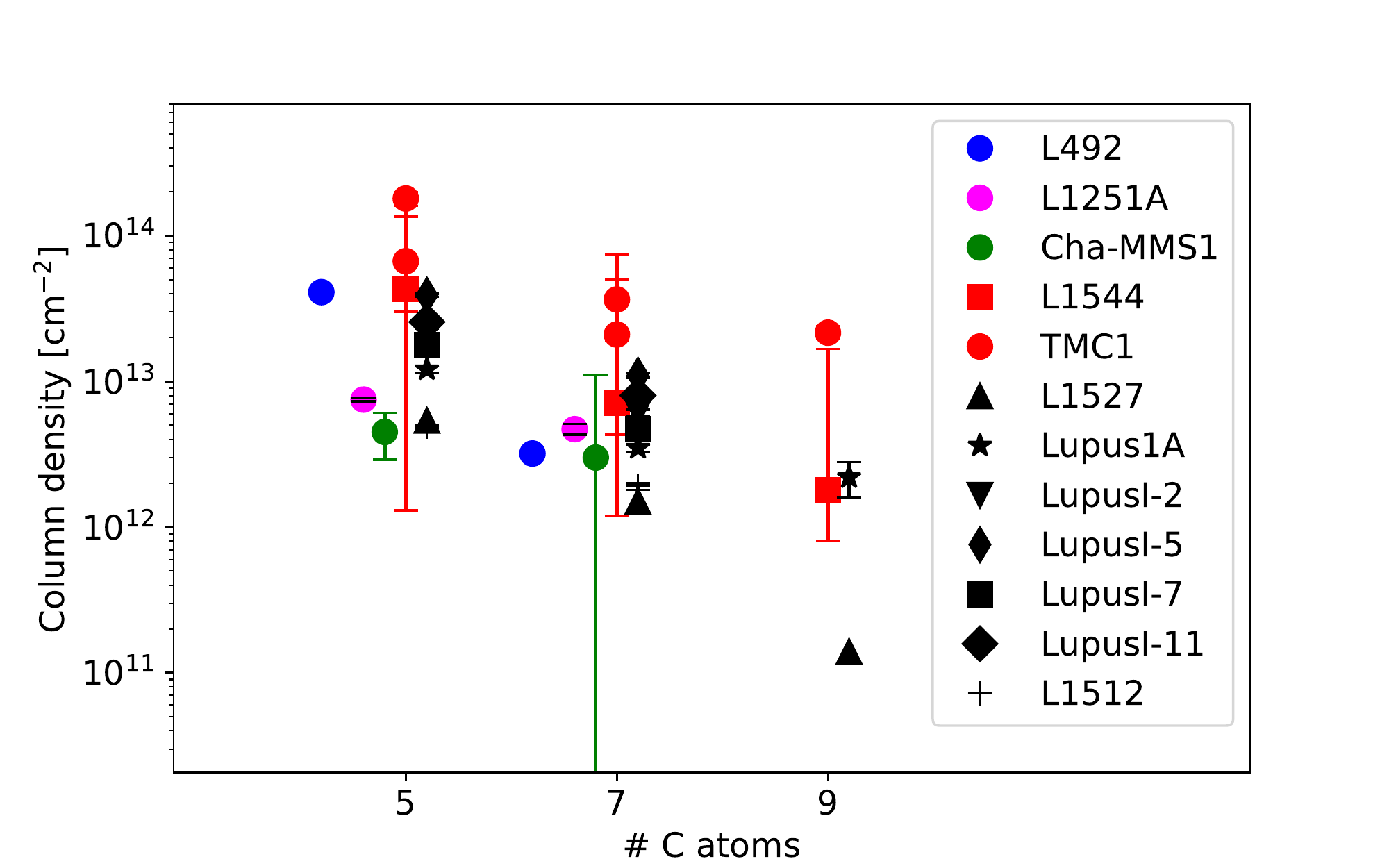}
    \includegraphics[width=1\columnwidth]{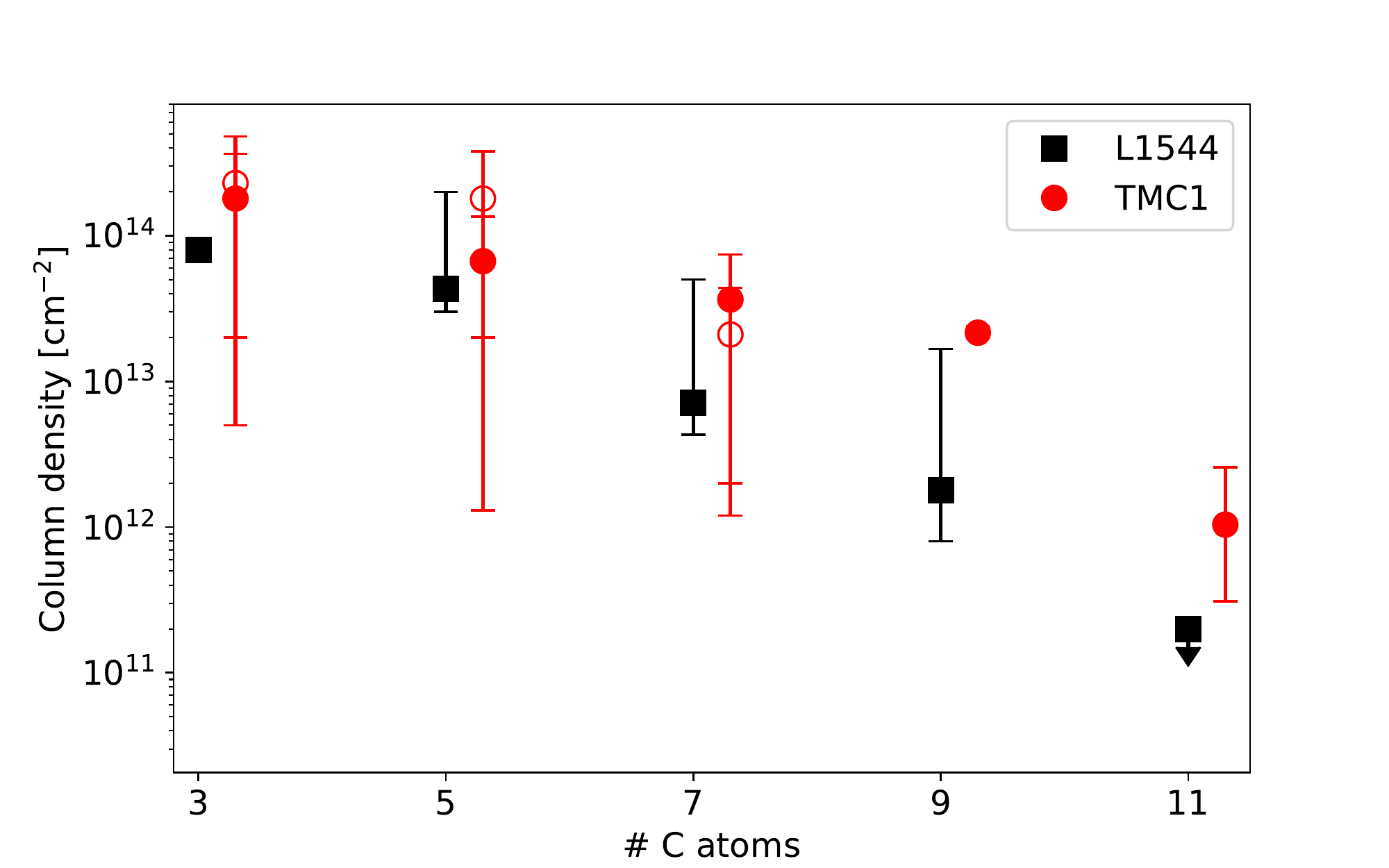}

    \caption{Comparison between the cyanopolyyne column densities measured in different star forming regions.
    \textit{Left panel}: Comparison of column densities of HC$_5$N, HC$_7$N and HC$_9$N. We indicate with the same color the cold cores located in the same star forming regions. The cyanopolyyne emission is commonly detected in several star forming regions and the column densities are in agreement within one order of magnitude, considering the errors and the different instruments. The measurements are taken from the references in Table \ref{Tab:cyano-other-sources}.
    \textit{Right panel}: Comparison between the cyanopolyyne abundances from HC$_3$N to HC$_{11}$N measured in TMC-1 and L1544 (this work). For TMC-1, the filled circles are from \citet{Loomis2021}, while the empty ones refer to \citet{Cernicharo2020} and \citet{Cabezas2022}.}
    \label{fig:cyano-comparison}
\end{figure*}

Cyanopolyyne emission is ubiquitous in the ISM. 
The simplest cyanopolyyne, HC$_3$N, was one of the firsts molecular species detected outside of our Galaxy \citep{Mauersberger1990}. 
Small cyanopolyynes, up to HC$_7$N, have been detected in several starless and protostellar cores in different star forming regions.
However, few measurements exist so far of more complex cyanopolyynes such as HC$_9$N and HC$_{11}$N. 
Figure \ref{fig:cyano-comparison} (left panel) and Table \ref{Tab:cyano-other-sources} report the column densities measured in starless cores for cyanopolyynes from HC$_5$N to HC$_9$N.
The only source for which all the cyanopolyynes from HC$_3$N to HC$_{11}$N have been measured so far is TMC-1.
Figure \ref{fig:cyano-comparison} (right panel) shows the comparison between L1544 (this work) and TMC-1 for cyanopolyynes from HC$_3$N to HC$_{11}$N. 
The deep surveys QUIJOTE and GOTHAM, performed on the source with the Yebes 40m telescope and the GBT, respectively, lead to precise column density measurements. The major uncertainties are introduced by the assumption on the source size, which is highly covariant with the column density.
Moreover, since the cyanopolyynes are very abundant in the source, the column densities of some species such as HC$_5$N and HC$_9$N could be affected by line opacity effects \citep[][and private communication]{Cernicharo2020}.
For all these reasons, every comparison between the column densities measured in different sources has to be taken with caution.
In TMC-1 all the cyanopolyynes have higher column densities than in L1544. In particular N(HC$_3$N) is higher of a factor 2--3, while N(HC$_5$N) of a factor 2--4, N(HC$_7$N) of a factor 3--5 and N(HC$_9$N) of a factor 12. The associated errors are quite large and further constraints on the cyanopolyyne spatial distribution are needed in order to derive reliable abundances measurements and to effectively constrain the formation routes.
However, the measurements in L1544 confirm that TMC-1 is not a unique source but an active carbon chains chemistry is efficient also in other sources. 
The comparison between the HC$_5$N, HC$_7$N and HC$_9$N column densities, reported in Figure \ref{fig:cyano-comparison} (left panel), suggest that the same chemistry could be active also in other cold cores. The fact that the measured column densities are similar within one order of magnitude, considering the different instruments and the large errors, may suggest that the heavy (with more than 5 carbons) cyanopolyyne formation is similar in different star forming regions, once free gaseous carbon is available (see Subsection \ref{subsec:discussion-tmc1}).

\begin{table*}[]
    \centering
    \begin{tabular}{cc|ccccc}
    \hline
         Star forming region & Source & T$_{ex}$ & N(HC$_5$N) & N(HC$_7$N) & N(HC$_9$N) & References \\
          &  & K & 10$^{12}$ cm$^{-2}$ & 10$^{12}$ cm$^{-2}$ &  10$^{12}$ cm$^{-2}$ & \\
         \hline
        Taurus & TMC-1 & 8 & 66.9 (1.3) & 36.5 (1.3)  & 22 (2)  & [1]\\
        Taurus & TMC-1 & 8.6 (0.2) -- 7.6 (0.2) & 180 (20) & 21 (2)  & --  & [2,3]\\
        Taurus & L1527 IRAS 04368+2557 & 12.3 & 5.4 & 1.5 & 0.14 & [4]\\
        Lupus & Lupus-1A & 10.0 (0.2) & 12.1 (0.6) & 3.5 (0.2) & 2.2 (0.6) & [5,6,7]\\
        Serpens & Serpens South 1a & 7 & 12 (1) & 6.0 (0.2)  & 3.1 (0.2)  & [8]\\
        Lupus & LupusI-2 & 11.5 (0.2) & 22 (1) & 6.1 (0.3) &  -- & [7]\\
        Lupus & LupusI-5 & 11.2 (0.1) & 39 (1) & 11.0 (0.4) &  -- & [7]\\
        Lupus & LupusI-7/8/9 &  10.2 (0.1) & 17.7 (0.6) & 4.7 (0.2) &  -- & [7]\\
        Lupus & LupusI-11 & 11.9 (0.8) & 25.6 (0.8) & 8.0 (0.3) &  -- & [7]\\
        Chameleon & Cha-MMS1 & 7 (1) &  4.5 (1.6) & 3 (8) & -- & [9]\\
        Taurus-Auriga & L1512 & 8.7 (0.7) &  4.9 (0.1) & 1.9 (0.1) & -- & [10]\\
        Cepheus & L1251A & 6.2 (0.3) &  7.5 (0.2) & 4.7 (0.4) & -- & [10]\\
        Aquila Rift & L492 & 6.5--10 & 41  & 3.2  & -- & [11]\\
         \hline
    \end{tabular}
    \caption{Cyanopolyynes abundances in starless cores. 
    [1] \citealt{Loomis2021}; 
    [2] \citealt{Cernicharo2020}; 
    [3] \citealt{Cabezas2022} ;
    [4] \citealt{Sakai2008}; 
    [5] \citealt{Sakai2009}; [6] \citealt{Sakai2010}; [7] \citealt{Wu2019};
    [8] \citealt{Li2016};  
    [9] \citealt{Cordiner2012}; [10] \citealt{Cordiner2011}; [11] \citealt{Hirota2006}.
    \label{Tab:cyano-other-sources}}
\end{table*}

\subsection{Chemistry of cyanopolyynes} \label{subsec:discussion-chemistry}

There is an ample consensus that the formation of cyanopolyynes is dominated by gas-phase reactions, in contrast with other large (e.g. with more than 5 atoms) species where a dust-grain surface chemistry can be at work \citep[e.g.][]{Ceccarelli2022}.
The reason is that unsaturated carbon chains, such as the cyanopolyynes, would rapidly be hydrogenated on the dust-grain surfaces so that, in order to produce the large amount of observed cyanopolyynes and to have them in the gas-phase at the low temperatures where they are observed, the grain-surface chemistry seems not a viable solution.

Several gas-phase formation routes have been invoked in the literature.
In general, for cyanopolyynes with more than five carbon atoms the following reactions are believed to be (the most) important \citep[for a review see e.g.][]{Fukuzawa1998}:

\begin{tabular}{clcl}
 1 & C$_{2n+2}$H + N & $\rightarrow$ & HC$_{2n+1}$N + C \\ 
 2 & C$_{2n}$H$_2$ + CN & $\rightarrow$ & HC$_{2n+1}$N + H \\ 
 3 & H$_3$C$_{2n+1}$N$^+$ + e$^+$ & $\rightarrow$ & HC$_{2n+1}$N + H$_2$ \\
 4 & H$_2$C$_{2n+1}$N$^+$ + e$^+$ & $\rightarrow$ & HC$_{2n+1}$N + H \\
\end{tabular}
where $n\geq2$. 

Destruction routes are dominated by reactions with ions such as e.g., C$^+$, H$^+$, H$_3^+$ and HCO$^+$.

\cite{Loison2014} published a critical and general review of the reactions involving carbon chains, including cyanopolyynes up to HC$_9$N.
They only list reactions (1), where they roughly evaluate the rate and branching ratios based on the capture theory and exothermicity of the products, respectively.
Since the rate constants of reactions (3) are considered the same for HC$_7$N and HC$_9$N, and the destruction also occurs at the same rate, the ratios HC$_5$N:HC$_7$N:HC$_9$N would reflect the parent species ratios, namely HC$_6$:HC$_8$:HC$_{10}$.
In other words, the HC$_n$N/HC$_{n+2}$N ratios is inherited from the HC$_n$/HC$_{n+2}$ one, as no reaction directly link HC$_n$N to HC$_{n+2}$N in the \cite{Loison2014} scheme.
Anyway, \cite{Loison2014} modeled the TMC-1 case and predicted the  HC$_5$N:HC$_7$N:HC$_9$N ratios to be 1:0.14--0.3:0.10--0.13 at around $1\times10^5$ yr, where the model predictions agree better with observations (of several carbon chains).
The two low and high value of each ratio are obtained assuming an elemental C/O abundance equal to 0.7 and 0.95, respectively.
In general, the HC$_7$N:HC$_9$N ratio observed toward TMC-1 is larger than the predicted one, even though the error bars are relatively large.
Likewise, the HC$_5$N:HC$_7$N:HC$_9$N ratios that we measured in L1544, about 1:6:4 (see Fig. \ref{fig:nonLTE-analysis}) are not consistent with the \cite{Loison2014} model predictions.

Reactions (2) was first mentioned and theoretically studied, via ab initio calculations, by \cite{Fukuzawa1998} for $n$=1--4.
These authors found that the reactions (2) with $n\geq2$ are exothermic and the transition state barriers are all embedded.
However, to our best knowledge, \cite{Fukuzawa1998} did not obtain the kinetics of the reactions.
Reactions (3) and (4) were first introduced by \cite{Herbst1989} and successively studied by \cite{Loomis2016}.
These authors obtained roughly estimates of their rate constants by educated guess (extrapolation from similar reactions with small carbon chains or via the Langevin rate).
However, again to our best knowledge, no specific experimental or theoretical ab initio studies exist in the literature on the rate constants and branching ratios of reactions (3) and (4).
That said, \cite{Loomis2016} developed an astrochemical model to reproduce the observations of TMC-1 and predicted HC$_5$N:HC$_7$N:HC$_9$N ratios equal to about 1:4:4, which is in relatively good agreement with our measured ratios towards L1544.
Finally, comparing observations of the $^{12}$C/$^{13}$C cyanopolyynes towards TMC-1 and model predictions, \citet{Burkhardt2018} found that reaction (4) best reproduce the observations.
Other processes could account for the formation of HC$_7$N and HC$_9$N. For instance, the reaction C$_3$N + C$_2$H$_2$ was proved to be very fast in CRESU experiments also at very low temperature \citep{Fournier2014}. Therefore, the reactions of the C$_3$N radical with C$_4$H$_2$ and C$_6$H$_2$ are expected to be at least as fast because of the increased dimension of the molecular partner. In addition, reactions of C$_4$H (detected in this object) with HC$_3$N or HC$_5$N are also expected to be very fast as the reaction of C$_4$H with acetylene is in the gas-kinetics limit at very low T \citep{Bertheloite2010}. 

\subsection{Age or UV illumination?} \label{subsec:discussion-tmc1}

As already mentioned above, the crucial point is of the cyanopolyyne chemistry is the presence of gaseous carbon atoms, which, under standard conditions, are predominantly locked in CO molecules.
Two general cases are possible:
(a) either the object is very young and the locking of carbon atoms into CO is not complete yet or (b) carbon atoms are liberated from CO thanks to processes such as intense UV illumination or cosmic-ray irradiation, able to destroy a fraction of CO.
The first case has been advocated for TMC-1 \citep[e.g.][]{agundez2013}.
Following the discussion of Sec. \ref{subsec:discussion-origin}, the second case, specifically the intense UV illumination, seems to apply to L1544 \citep[e.g.][]{Spezzano2016,Spezzano2017}.

In the case of UV illumination, the column density of the region where carbon is atomic is dictated by the penetration of the UV field, which is approximately given by a dust column density with a visual extinction of about 2--3 mag \citep[e.g.][]{Hollenbach1997,Snow2006}, which corresponds to N(H$_2$)$\sim$4--6$\times 10^{21}$ cm$^{-2}$.
For example, the abundance of HC$_7$N (with respect to H$_2$) in L1544 would be $\sim 2\times10^{-9}$, where we used N(HC$_7$N)$\sim 8\times 10^{12}$ cm$^{-2}$ (\S~\ref{subsec:model-LVG}).

In the case of a very young cloud, the column density of cyanopolyynes is, instead, determined by the H$_2$ column density of the cloud itself.
For the TMC-1 cloud, the latest estimates indicate N(H$_2$) $\sim 1.5 \times 10^{22}$ cm$^{-2}$, which correspond to a visual extinction of about 15 mag  \citep{Fuente2019}.
Assuming N(HC$_7$N)$\sim 2\times 10^{13}$ cm$^{-2}$ (Sec. \ref{subsec:discussion-comparison}), the HC$_7$N abundance in TMC-1 is $\sim 1\times10^{-9}$, similar to L1544.

Therefore, based on the simple estimates of the $n\geq 5$ cyanopolyyne abundances of TMC-1 and L1544, we confirm the general affirmation that the cyanopolyyne chemistry only depends on the gaseous carbon abundance.
In other words, the $n\geq 5$ cyanopolyyne abundance ratios are the same regardless the cause of the presence of gaseous carbon atoms.
On the other hand, the $n\geq 5$ cyanopolyyne column density is a proxy of the cause of the gaseous carbon.
For example, a column density of HC$_7$N $\sim 8\times 10^{12}$ cm$^{-2}$ would be a strong indication of a UV-illuminated gas.

\section{Summary and Conclusions} \label{sec:conclusions}
We performed new observations using the GBT towards the prestellar core L1544. We detected several emission lines from cyanopolyynes from HC$_3$N to HC$_9$N, detected for the first time towards the source.
The resolved spectral profiles show a double-peak profile, suggesting that the bulk of the cyanopolyyne emission is associated to the southern region of the core, where other smaller carbon chains peak. This supports the idea that cyanopolyynes are mainly formed in the external part of the core, where the interstellar field radiation increases the free carbon atoms available to form long chains.  
We perform a large velocity gradient analysis of the observed HC$_5$N, HC$_7$N, and HC$_9$N lines, thanks to a new estimation of the collisional coefficients. The simultaneous fitting of the species, indicates a gas kinetic temperature of 5--12 K, a source size of 80$\arcsec$ and a gas density larger than 100 cm$^{-3}$. The HC$_5$N/HC$_7$N/HC$_9$N abundance ratios measured in L1544 are about 1/6/4. The measured column densities are lower by a factor 2 to 5 than those measured in TMC-1. Even if the measurements in other star forming regions are scarce, the results obtained in L1544 suggest that a complex C-chain chemistry is active in other sources and it is related to the presence of free gaseous carbon. The latter can be abundant either because the core is very young and the conversion to CO is not completed, or because the CO is destroyed by UV illumination or cosmic-ray irradiation. We suggest that the column density of heavy cyanopolyynes (larger that HC$_5$N) could be a proxy to discriminate between these two regimes.



\begin{acknowledgments}
We are grateful to Jos\'e Cernicharo for valuable discussions and suggestions. We thank the anonymous referee for the constructive comments.
This project has received funding from: 1) the European Research Council (ERC) under the European Union's Horizon 2020 research and innovation program, for the Project “The Dawn of Organic Chemistry” (DOC), grant agreement No 741002; 2) the PRIN-INAF 2016 The Cradle of Life - GENESIS-SKA (General Conditions in Early Planetary Systems for the rise of life with SKA); 3) the European Union’s Horizon 2020 research and innovation programs under projects “Astro-Chemistry Origins” (ACO), Grant No 811312; 4) the German Research Foundation (DFG) as part of the Excellence Strategy of the federal and state governments - EXC 2094 - 390783311. The Green Bank Observatory is a facility of the National Science Foundation operated under cooperative agreement by Associated Universities, Inc.\\
\end{acknowledgments}


\vspace{5mm}
\facilities{GBT}


\software{astropy \citep{Astropy1,Astropy2}, matplotlib \citep{Matplotlib}}






\bibliography{GBT-Main-text}{}
\bibliographystyle{aasjournal}
\end{document}